\documentclass[conference]{IEEEtran}
\IEEEoverridecommandlockouts
\usepackage{cite}
\usepackage{amsmath,amssymb,amsfonts}
\usepackage{algorithmic}
\usepackage{graphicx}
\usepackage{textcomp}
\usepackage[utf8]{inputenc}
\usepackage{graphicx}
\usepackage{textcomp}
\usepackage{indentfirst}
\usepackage{xcolor}
\usepackage{algorithmic}
\usepackage{amsmath,amssymb,amsfonts}
\usepackage[ruled,vlined]{algorithm2e}
\usepackage{graphicx}
\usepackage[export]{adjustbox}
\usepackage[left=2cm, right=2cm, top=2cm, bottom=2.413cm]{geometry}
\usepackage{xcolor}
\def\BibTeX{{\rm B\kern-.05em{\sc i\kern-.025em b}\kern-.08em
    T\kern-.1667em\lower.7ex\hbox{E}\kern-.125emX}}
\usepackage{comment}
\usepackage{dblfloatfix}
\begin{document}

\title{ 
Vulnerability Characterization and  Privacy Quantification for Cyber-Physical Systems
}



    

\author{
\IEEEauthorblockN{Arpan Bhattacharjee\IEEEauthorrefmark{1},
Shahriar Badsha\IEEEauthorrefmark{2},
Md Tamjid Hossain\IEEEauthorrefmark{3}, Charalambos Konstantinou\IEEEauthorrefmark{4}, Xueping Liang\IEEEauthorrefmark{5}
}
\IEEEauthorblockA{University of Nevada, Reno, NV\IEEEauthorrefmark{1}\IEEEauthorrefmark{2}\IEEEauthorrefmark{3}, KAUST, Saudi Arabia\IEEEauthorrefmark{4}, University of North Carolina, Greensboro, NC\IEEEauthorrefmark{5}\\
\IEEEauthorrefmark{1}\IEEEauthorrefmark{3}\{abhattacharjee,mdtamjidh\}@nevada.unr.edu,
\IEEEauthorrefmark{2}sbadsha@unr.edu,
\IEEEauthorrefmark{4}charalambos.konstantinou@kaust.edu.sa,\\
\IEEEauthorrefmark{5}x.liang@uncg.edu
}}

\maketitle

\begin{abstract}

Cyber-physical systems (CPS) data privacy protection during sharing, aggregating, and publishing is a challenging problem.
Several privacy protection mechanisms have been developed in the literature to protect sensitive data from adversarial analysis and eliminate the risk of re-identifying the original properties of shared data. However, most of the existing solutions have drawbacks, such as (i) lack of a proper vulnerability characterization model to accurately identify where privacy is needed, (ii) ignoring data providers privacy preference, (iii) using uniform privacy protection which may create inadequate privacy for some provider while overprotecting others, and (iv) lack of a comprehensive privacy quantification model assuring data privacy-preservation. To address these issues, we propose a personalized privacy preference framework by characterizing and quantifying the CPS vulnerabilities as well as ensuring privacy. First, we introduce a Standard Vulnerability Profiling Library (SVPL)  by arranging the nodes of an energy-CPS from maximum to minimum vulnerable based on their privacy loss. Based on this model, we present our personalized privacy framework (PDP) in which Laplace noise is added based on the individual node's selected privacy preferences. Finally, combining these two proposed methods, we demonstrate that our privacy characterization and quantification model can attain better privacy preservation by eliminating the trade-off between privacy, utility, and risk of losing information.
\end{abstract}

\begin{IEEEkeywords}
Cyber-physical systems (CPS), Privacy, Vulnerabilities, Differential Privacy (DP), Personalized Differential Privacy.
\end{IEEEkeywords}

\vspace{-3mm}
\section{Introduction}
\noindent Cyber-physical systems (CPS) such as manufacturing plants and smart grids generate, transmit, and collect a tremendous amount of privacy-sensitive data. This data ensures a wide range of advantages to all involved system's parties but also raises a privacy risk to data providers whose shared data can be statistically correlated to identify private information \cite{giraldo2017security}.
Additionally, the lack of a comprehensive vulnerability assessment, robust data protection, and attack-defense mechanism drives adversaries to compromise CPS nodes\cite{lee2019security}. 
Topological investigations by researchers have shown that energy-CPS smart grids have inter-dependencies between their different nodes\cite{cetinay2018nodal}. Some of these nodes have relatively higher importance as they generate essential data for the grid operation. Identifying the vulnerabilities of these critical nodes and ensuring a robust privacy-preservation mechanism can improve the system's resiliency against privacy-compromising attacks \cite{aman2019data}. 
Privacy preservation of sensitive data (e.g., electricity usage data, utilities power generation data, or electrical appliances location data) carries paramount importance since their compromise can affect the secure and reliable operation of the grid. To counter this privacy problem, several solutions (e.g., encryption, differential privacy (DP) \cite{hossain2021privacy}, k-anonymity, l-diversity, etc.) have been proposed \cite{begum2018comparative,9562929,bhattacharjee2021integrity}. However, these solutions lack a proper privacy characterization and measurement technique to identify the expected privacy requirements of individual data providers.


Currently, uniform data privacy is provided to all the data providers of the smart grid infrastructure, leading to inadequate privacy preservation for some data providers while over-protecting others \cite{jorgensen2015conservative}. This uniform privacy protection also introduces an unacceptable amount of noise, resulting in poor data utility and system performance degradation. To eliminate privacy problems in CPS, and specifically in smart grids, our work focuses on designing a comprehensive CPS (power) data privacy characterization and quantification model to identify the vulnerabilities that occur during data sharing, publishing, or aggregating between the associated parties. This privacy
quantification model is subsequently integrated with the Personalized Differential Privacy (PDP) solution in which data providers specify their personal data’s privacy requirements. This integrated privacy characterization and protection model ensures comprehensive grid vulnerability assessment as well as data point diversity, and indistinguishability. As a result, adversaries who run malicious analytics on these private datasets will not be able to identify sensitive private data.

In this paper, we focus on introducing a vulnerability characterization and privacy quantification model for CPS. The contributions can be summarized as follows:

\begin{itemize}
    \item We introduce a Standard Vulnerability Profiling Library (SVPL) using an attack graph and networks centrality metrics to combine the risk sources, data privacy weaknesses, feared events, and associated harms to arrange nodes in a rank from minimum to maximum vulnerable based on their privacy loss scenario. 
    
    \item We develop a Personalized Differential Privacy (PDP) model where data privacy is guaranteed based on individual data providers' selected privacy preferences. 
    Our PDP model provides better data utility and privacy over traditional Uniform Differential Privacy (UDP).
    
\end{itemize}

The rest of this paper is assembled as follows: Section \ref{sec:review} provides an overview of the vulnerability characterization and privacy-protection research in literature. The formulation of the Standard Vulnerability Profiling Library (SVPL) module is demonstrated step by step in section \ref{sec:Vul_Chac_Mod}. Section \ref{sec:Proposed} describes the formulation of a PDP model where privacy is dependent on the individual node's privacy preference. Section  \ref{sec:Experiment} validates the proposal using experimental analysis. Finally, Section \ref{sec:Conclusion}  concludes the paper. 

\section{Related Work} \label{sec:review}
\noindent Due to the integration of CPS in critical infrastructures and the increasing number of attack incidents in such interoperable and interconnected architectures \cite{peterson2010ics}, characterizing and quantifying the vulnerabilities of CPS is of paramount importance. Researchers have proposed vulnerability analysis based on attack graphs\cite{gonda2018analysis}, standard vulnerability scoring systems (CVSS)\cite{mavroeidis2017cyber}, and network centrality metrics \cite{peng2020security}. However, defense mechanisms such as deployments of firewall or intrusion detection systems are proved ineffective \cite{yaacoub2020cyber} to protect the CPS. Moreover, the lack of proper vulnerability characterization models also makes it difficult to accurately identify the system's vulnerabilities and provide an appropriate situational awareness capability.


Nowadays, data privacy preservation has become an issue of paramount importance. Aggrawal et al. \cite{aggarwal2008general} first introduced data privacy preservation in 2008 to mitigate privacy-compromising access and analytics to industrial system's private data.
The goal of privacy-preserving techniques is to protect private information against being published or revealed by unauthorized as well as authorized users \cite{keshk2019integrated}. Several privacy-preserving approaches such as encryption-based, perturbation-based, authentication-based, and technologies like blockchain have been proposed in the literature to protect the sensitive information of CPS effectively \cite{pan2019big, bhattacharjee2020block, hossain2020porch, 9239055, 9291600}. Regarding privacy preservation works in CPS, Gope et al.\cite{gope2019efficient} proposed a privacy-preserving scheme to analyze the energy consumption data of end-users securely. Other researchers like Gai et al.\cite{gai2019privacy} have introduced a blockchain-based technique to address privacy leakage challenges in smart grids.
 
There are some existing studies on differential privacy (DP) and Laplace noise addition mechanism in the existing literature to accomplish effective data privacy protection \cite{keshk2019privacy, 8854247, hossain2021desmp}. However, not many researchers have focused on personalized differential privacy (PDP) least of PDP in CPS to improve the data utility while preserving CPS data privacy. Recently, PDP-based techniques have become popular in social networks \cite{gong2015personalized}, location-based applications \cite{9443990}, and recommendation systems \cite{9527140} because of their optimization power to improve data utility while assuring privacy. To counter these limitations of existing works, we have adopted PDP for the CPS system by characterizing the system's vulnerabilities, which will help achieve optimal data privacy while maintaining utility and reducing the risk of data disclosure.
 


\section{Proposed Vulnerability Characterization
} \label{sec:Vul_Chac_Mod}

\subsection{Smart Grid Topological Relationship}
\noindent Electric grids can be represented with nodes (system buses) and their associated connectivity links (e.g., transmission and distribution lines). Such relationship can be modelled as a graph $G(\mathcal{N}, \mathcal{L})$ where $\mathcal{N}$ and $\mathcal{L}$ represents the number of nodes and communication links. Multiple communication links connected to the same node pair are considered as one link here. The interconnectivity of the graph $G(\mathcal{N}, \mathcal{L})$ is represented by a $N \times N$ adjoint matrix $\mathcal{A}$ where $\mathcal{A}_{i k}=1$ if the nodes $i$ to $k$ are coupled by a direct link; otherwise $\mathcal{A}_{i k}=0 .$ The $N \times N$ adjacent matrix is presented as a Laplacian matrix $L$ to present the properties of the grid topology in $\mathbf{L}$ = $\Delta$ - $\mathcal{A}$, where $\Delta=\operatorname{diag}\left(e_{1}, \ldots, e_{N}\right)$ is the diagonal edges of the matrix connecting the centers of opposite nodes with the edges and their associated elements $E_{i}=\sum_{k=1}^{N} a_{i k}$.
\vspace{-1mm}
\subsection{Vulnerability Characterization Framework}
 \begin{figure} 
\centering
\includegraphics[width=0.99\linewidth]{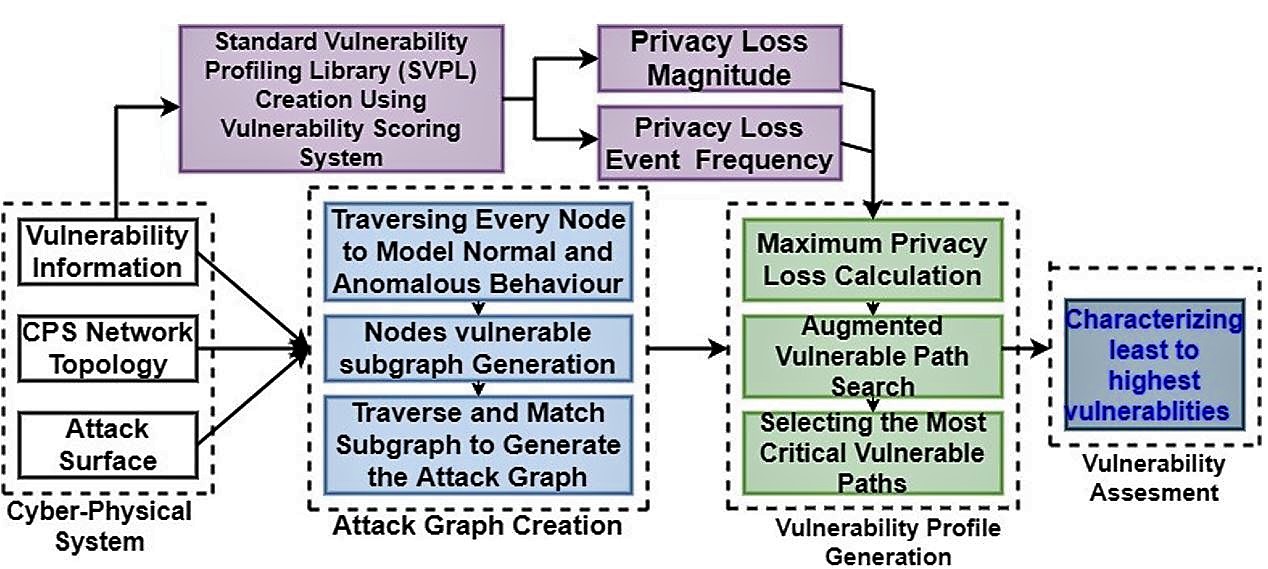} 
\caption{Vulnerability Characterization Framework.} 
\label{fig: Vulnerability}
\vspace{-10pt}
\end{figure}
\noindent In this work, we use an attribute-based attack graph to assess the network's vulnerabilities. The vulnerability characterization framework incorporates five parts as shown in Fig. \ref{fig: Vulnerability}. That are CPS network properties, attack graph creation, vulnerability profile generation,  privacy loss assessment using SVPL and maximum privacy loss calculation, and at last vulnerability characterization and quantification module. The evaluation includes the following steps:
 
 \begin{itemize}
     \item First, we generate an attack graph by analyzing existing information like the grid's topological relationship to traverse and assess the previous attacked incidents to collect the vulnerabilities of the compromised nodes.  
     \item This attack graph is then leveraged to build SVPL by calculating the vulnerability scores generated from risk sources, data privacy weaknesses, feared events, and associated harms. 
     
    \item  According to the computed vulnerability score, the maximum to minimum privacy loss is calculated by combining the privacy loss magnitude and the frequency of the successful privacy-compromising attacks.
\item Finally, this maximum privacy loss assists in identifying the network's key fragile points by creating the best attack profile that will help the grid utilities to reinforce their budgets appropriately and ensure defense before an attack occurs. 
 \end{itemize}
 
 
\subsection{Vulnerability Assessment by Building an Attack Profile}
\noindent In order to perform a complete privacy risk assessment and develop the vulnerability attack profile, we need at first to establish a common link by defining the relationship between the ``feared events", ``risk sources", and ``privacy weaknesses" to model their impacts called ``privacy harm".

\textbf{Definition 1} \textit{(Risk Source): Risk sources define an individual or organizational entities whose legitimate or illegitimate owning of a dataset and actions causes intentional or unintentional privacy harm to the system.}

\textbf{Definition 2} \textit{(Privacy Weaknesses): Privacy weakness is a limitation of the data protection mechanisms' security measures, including poor choice of security functionality or error-prone system design, which ultimately causes privacy harms in the system.}
 
\textbf{Definition 3} \textit{(Feared Event): Feared events are the malicious actions on the system that exploit the system's privacy weaknesses, leading to privacy harm. }

\textbf{Definition 4} \textit{(Privacy Harm): Privacy harm is the negative impact on datasets, individual data providers, or the network resulting from one or more feared events to compromise the system's physical, cyber or financial stability. }

 Based on these four attributes and traditional attack graph model\cite{liu2016network}, we develop our vulnerability assessment attack graph (VAAG), which comprises eight tuples as follows: 
 $VAAG$ = $(N, E, PC, M, D, R, PLM, FPLE)$

\begin{itemize}
    \item $N$ represents the nodes of the system which consists of start node $S_{n}$, attacked nodes $A_{n}$ as well as attacked path $A_{p},$  where, $N=$ $S_{n} \cup A_{n} \cup A_{p} .$ If $A_{n} \in A_{p},$ then $A_{n}$ is a compromised node on the attack path.
    
\item $E$ represents the edge set of all $N$ nodes where $E \subset\left(S_{n} \times A_{p}\right) \rightarrow$ $\left(A_{p} \times A_{n}\right),$ and where every privacy attack is denoted by an action  $a$ in $E$. 

\item $P C$ denotes the physical connectivity of the nodes, where $ P C \subseteq\left(N \times N_{l}\right)$. $N_{l}$ is the physical link between the nodes.

\item $M$ is a mapping function that maps the edge sets $E$ to the compromised set $C$, which is $M: E \rightarrow C$. In the VAAG, every malicious attack $(a \in E)$ is the result of a vulnerability $C(e) $ on that node according to two privacy risk factors: Privacy Loss Magnitude (PLM) and Frequency of Privacy Loss Events (FPLE).

\item $D$ denotes the vulnerability dependencies of the nodes presented by $d_{i}: v_{i} \rightarrow v_{j},$. This means if adversaries try to compromise node $N$ then both $v_{i}$  and $v_{j}$ vulnerability set is needed. 

\item The privacy risk score is calculated by decomposing two risk factors: PLM  which denotes the privacy attacks on the data points; and FPLE, which presents the successful occurrences of an attack by a malicious actor. This PLM and FPLE are calculated from the existing vulnerability information of the CPS infrastructure. The overall risk is calculated by combining these risk factors: $R$ = $P L M$ * $F P L E$. 

\end{itemize}
 
\subsection{Standard Vulnerability Profile Generation}
\noindent The formulation of our proposed SVPL to identify and characterize the vulnerabilities of the CPS nodes mainly consists of three parts: traversing the compromised nodes, developing an attack graph, and using these sub-level attack graphs to generate the nodes' global attack profile. In this process of traversing and identifying the compromised nodes, we start our process by going through the existing vulnerability information of the nodes $V$ and the attack incidents $A I$ and using it as a baseline profile to generate global vulnerability assessment attack graph ``VAAG''. The attack incident set $A I$ is the set of attack preconditions $A I _P$ and attack consequences $AI _ C$. The vulnerability profile generation follows Algorithm 1,  where $A I$ at first checks if there is an available attack incident by traversing and matching the current information with the system's existing information. If abnormalities are found in the new node $N$ then it is considered as a compromised node. This compromised node $ N$ will be used to find the previously connected nodes by linking through the edges and putting the whole compromised node-set in the $C N$ and their edges in $E$. If this newly vulnerable node is not included in the $V$ set, it will be added to the existing vulnerable node set.  Finally, every node of $C N$ will be matched with the $ AI$ set to profile the different vulnerabilities. 

\begin{algorithm}
\caption{Vulnerability Profile Generation}
\textbf{Input:} Nodes existing vulnerability set $V$, attack incident set $AI$ and newly added nodes $N$.

\textbf{Output:} Newly Compromised nodes $CN$ and their edge sets $E$. 

\textbf{Step 1:} if $A I \neq \emptyset$ then \textbf{Return}


\textbf{Step 2:} for each $a i \in A I$

\textbf{if} $\mathrm{N} \in a i \cdot$ $\operatorname{AI_p} \wedge \left(a i \cdot AI_p-\{ N \}\right) \subset$
V ; \textbf{then}

\hspace{1 cm}1. Start attack node creation $A_n$

\hspace{1 cm}2. $ C N \leftarrow CN \cup\{A_n\}$

\hspace{1 cm}3. for each targeted nodes $T_n \in a i \cdot AI_p$

\hspace{1 cm}4. $\quad E \leftarrow E \cup\{<T_n, A_n>\}$

\hspace{1 cm}5. for each $T_n \in a i \cdot AI_c$

\hspace{1 cm}6.\hspace{.40 cm} if $T_n \notin$ V then

\hspace{1 cm}7. V $\leftarrow$ V $\cup\{$ 
$T_n$\} $\}$; \textbf{End}

\textbf{Step 3:} for each $A_n \leftarrow$ N 

\hspace{1 cm}8. CN Match $( V, A I, AI_p$ \& $AI_c )$

\hspace{1 cm}\textbf{End}; \textbf{Return} $CN , E$

\end{algorithm}

\vspace{-5mm}
\section{Personalized Differential Privacy} \label{sec:Proposed}
\noindent This section demonstrates our proposed PDP solution based on traditional differential privacy. We assume every source node is a data provider who owns the system's private data. Similarly, any destination node (fog, cloud) is involved in data transmission and aggregation. This paper aims to ensure every source node will share their sensitive data with the destination nodes under a PDP protection mechanism. Our privacy characterization module provides an SVPL to identify the adversaries and vulnerabilities beforehand but can not guarantee privacy preservation of raw data in real-time. To eliminate this limitation, our proposed personalized privacy protection model 
ensures privacy by adding customized noise generated from Laplace distribution which depends on every source node's preference that is shown in Table \ref{tab:sec_analysis}. For example, based on the criteria of access to nodes distance is crucial because longer distances (more hops) will force the transmitting data to travel through more malicious nodes and raises the likelihood of getting compromised.  To measure this distance, we follow network centrality metrics (e.g., degree, eigenvector, betweenness, and closeness centrality) to rank the nodes in the graph and formulate the weighted smart grid architecture and employ the shortest path algorithm to find the actual distances between the nodes. 


\subsection{Personalized Differential Privacy Modelling}
\noindent In this section, we modeled the nodes (source, destination) based on their privacy preference.
We assume that a source node $S_{n}$ when transferring its private data $p_{d} \in P_D,$ where $P_D$ presents the private data needs privacy protection before sending the data to the destination. The privacy protection mechanism starts with the source node $S_{n}$ when releasing the data to its adjacent fog nodes $F_{n}$; it at first calculates a approximate data $d_{s}^{\prime}$ by mixing different noise with the original data and release this $d_{s}^{\prime}$ data instead of the sources original data $P_D$. This approximate data $d_{s}^{\prime}$ needs to satisfy the source nodes privacy protection preference $\varepsilon \in\left(\frac{1}{\Delta_{i j}}\right)$ where $\Delta_{i j}$ is the distance function $D: \Delta_{i j}: S_n \rightarrow D$ and $\varepsilon: D^{\kappa} \rightarrow D^{\kappa}$  presents the mapping function that transforms the distance $\Delta_{i j}$ into $\varepsilon \in\left(\frac{1}{\Delta_{i j}}\right)$ personalized  differential privacy preference level.

\subsection{Calculating Distance between Nodes of the Graph}
\noindent In this layered grid architecture, attackers can compromise private data from any source/destination node. This means the more significant transmission distance between the source and destination node causes more nodes to be involved in each one-hop distance for receiving, storing, and sending the transmitted data. This causes more untrusted nodes to be involved, raising the risk of getting compromised by malicious adversarial attacks. However, to keep the privacy protection approach stable and efficient, we set a maximum threshold distance $TH_{d}$ beyond which we will provide a random privacy loss ($\varepsilon$). So, the nodes' effective distance calculation in our weighted graph follows a least-cost approach using the ``Dijkstra Algorithm'' where the shortest path is calculated by finding the least cost from one node to every other connected node. So, the distance  $\Delta_{i j}$ between the source and destination node is calculated as follows:
$\Delta_{i j} = D(n) - S(n)$.

\subsection{Laplace Distribution and $\varepsilon$-Personalized Privacy}
\noindent The Laplace mechanism adds preferred noise to the private data from the Laplace distribution to fulfill each node's personal differential privacy preference. The privacy loss is employed by using the Laplace mechanism $L$ to inject $\varepsilon$-personalized private noise $\operatorname{Lap}(\frac{\Delta f}{\varepsilon})$ to the sensitive data $D$, that can be denoted as:
$L = f(D) + \operatorname{Lap}(\frac{\Delta f}{\varepsilon})$. This ensures $\varepsilon$-differential privacy to transform each nodes' private data $P_D$ to a noisy data $d_{s}^{\prime}$. 



















 
 
 
 
 

\begin{table}
   \centering
    \caption{Criteria Mapping to Personalized Privacy Preservation.}
    \begin{tabular}{|p{4.00cm}|p{.70cm}|p{.90cm}|p{.70cm}|}
    \hline
      \textbf{Criteria/Nodes} & \textbf{Edge} & \textbf{Fog} & \textbf{Cloud}\\
      \hline
      Access to nodes & High & Medium & Low\\
      \hline
      Access frequency to Private Data & High & Medium & Low\\
      \hline
      Attackers background knowledge & Low & Medium & High\\
      \hline
      Communication medium & Low & Medium & High\\
      \hline
      Operational complexity & High & Medium & Low\\
      \hline
    \end{tabular}
    
    \label{tab:sec_analysis}
\end{table}
\subsection{Mechanism of Achieving PDP}
\noindent The formulation of PDP follows sequential composition mechanism which guarantees the composition of $\sum_{i}^{l} \varepsilon_{i}$. For example, there is a set of source nodes  $S=$ $\left\{S_{1}, S_{2} \ldots S_{n}\right\},$ and their corresponding Laplacian mechanisms $L=$ $\left\{L_{1}, L_{2} \ldots L_{m}\right\},$  generates $\varepsilon_{1}, \varepsilon_{2} \ldots \varepsilon_{m}$ privacy. For every $L_{i}$ needs ensuring $\varepsilon$-differential privacy which ultimately guarantees the composition of $\sum_{i}^{l} \varepsilon_{i}$-differential privacy. Here, $\sum_{i}^{l} \varepsilon_{i}$ refers upper privacy bound of sequential composition mechanism. So, when a dataset is released from a node it goes through randomized privacy loss ($\varepsilon$) based on individual nodes preference and the final privacy guarantee will be the summation of total privacy loss $\varepsilon$.

\subsection{Criteria Mapping to Personalized Privacy Preference} \label{sec:criteria}
\noindent The traditional differential privacy follows a uniform privacy protection for all nodes; thus, it compromises the data utility. To overcome this limitation, we have adopted a privacy preference profile that is assigned to all of the nodes based on some criterion as shown in Table \ref{tab:sec_analysis}. The privacy preference of each node is mapped as high, medium, and low depending on the desired privacy and utility requirement. For example, it is comparatively easier to access the edge nodes than the fog and cloud nodes. Hence, privacy and security attacks (e.g., eavesdropping, membership inference, data poisoning, etc.) are more likely to occur in the edge nodes than the other nodes which is why high data privacy is desired in the edge nodes. Here to find the appropriate $\varepsilon$-differential privacy, we use the following mapping function:
$L\left(\varepsilon_{i}\right)=\frac{\exp \left(K_{i}^{t} x\right)}{\sum_{m=1}^{M} \exp ^{K M}}$;
where,  $i=1$, $K$ presents a parameter, and $K_{M}$ is the mapping parameter.


\begin{figure*}[b!]
\centering
\includegraphics[width=0.99\linewidth]{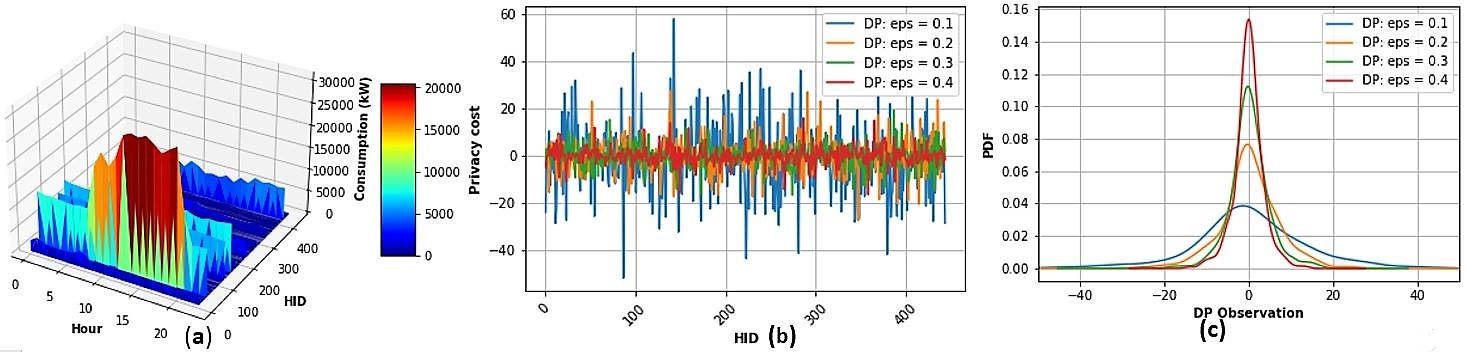} 
\caption{DP Evaluation: (a) Actual Consumption, (b) Impact of DP over actual consumption with varying privacy Loss, and (c) Distribution of Noise.} 
\label{fig: Personalized_DP}
\vspace{-10pt}
\end{figure*}

\subsection{Proof: Personalized Differential Privacy (PDP)}
\noindent To guarantee personalized $\varepsilon$-privacy, we at first take a source node with two different privacy preference $\varepsilon_{p}$ and $\varepsilon_{p+1}$, where $\varepsilon_{p+1}>\varepsilon_{p}$ and similarly we take another node with two different privacy level $\varepsilon_{q}$ and $\varepsilon_{q+1},$ where $\varepsilon_{q+1}>\varepsilon_{q}$. The sequential composition mechanism of our system is built upon the upper bound which guarantees $\sum_{i}^{n} \varepsilon_{i}$ privacy. So, the PDP for the two nodes is as follows: $P D P $ ($\varepsilon_{p}$ + $\varepsilon_{q}$) = $P D P $ ($\varepsilon_{p}$ + 1) or $P D P $ ($\varepsilon_{p}$ + $\varepsilon_{q}$) = $P D P $ ($\varepsilon_{q}$ + 1). 
Upon observing this equation, we can see $\varepsilon_{p}<\varepsilon_{q+1}$ and $\varepsilon_{q}<\varepsilon_{p+1}$ which refers that the two nodes with noisy responses generate different level of privacy-preserved data.

\section{Experimental Analysis} \label{sec:Experiment}
\noindent We conduct the model evaluation based on following criteria:

\subsection{ Utility Analysis} 
\noindent In the CPS infrastructure, the data is generated from the edge nodes and transmitted to the cloud node through the fog nodes. In edge nodes, the data utility is `1' and data privacy is `0' since edge nodes generate and preserve granular level data where no privacy is added to the data. Privacy is added only when edge nodes start sharing their data with the fog and cloud nodes. Any privacy-preserving mechanism (e.g., differential privacy) incurs utility loss over the original value which can be measured through different methods (e.g., MSE, MAE, Max\_Error, RMSE, MedAE, etc.). In our experiment, we interpret the utility loss of the fog and cloud nodes by Mean Absolute Error (MAE) which can be calculated using the following formula:
$\operatorname{MAE}$ = $\frac{\sum_{i=1}^{n}\left|y_{i}-x_{i}\right|}{n}$ where, MAE = mean absolute error; $y_{i}$= privacy preserved value; $x_{i}$= original value and 
$n$ = total number of data points. To measure the data utility loss on the scale of `$0$' to `$1$', we have calculated the percentage deviation (PD) using MAE and Mean  (i.e., $PD = MAE/Mean$) and then subtracting the percentage deviation from `$1$'. Here, the data utility of `$1$' reflects the highest possible utility of the nodes when no privacy preservation mechanism is applied and vice versa.

\begin{figure*}
\centering
\includegraphics[width=0.99\linewidth]{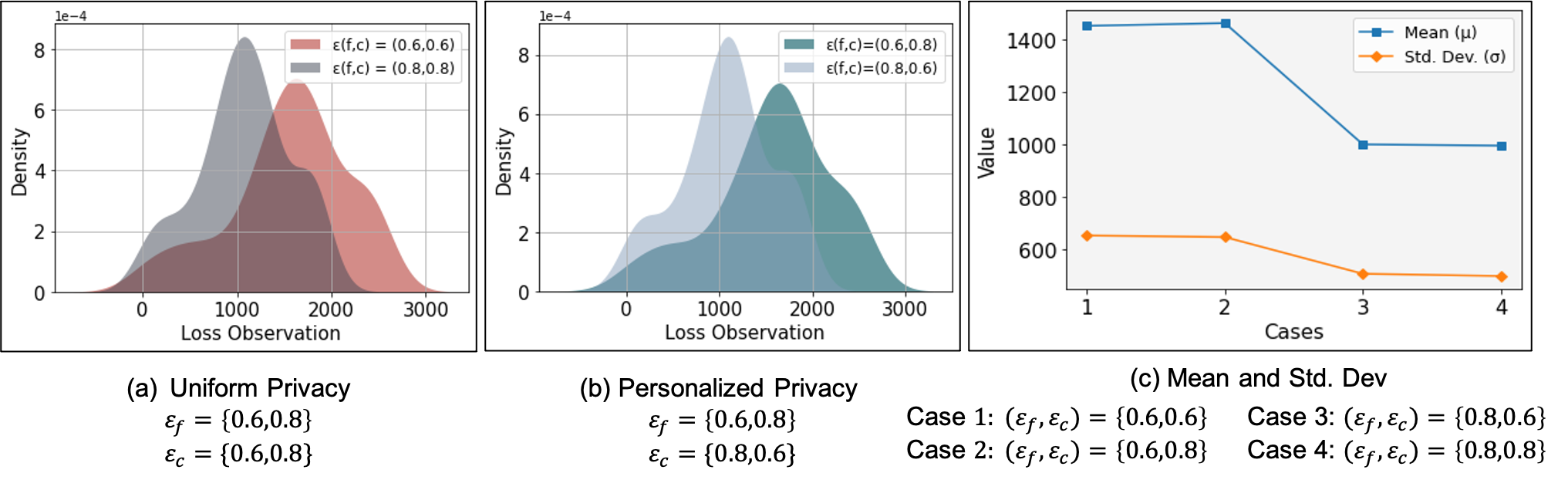} 
\caption{Distribution of losses over uniform and personalized privacy.} 
\label{fig: loss_DP}
\vspace{-10pt}
\end{figure*}

\subsection{ Privacy Analysis}
\noindent Our privacy analysis is conducted by generating an interactive privacy preference profile based on some criteria as shown in Table \ref{tab:sec_analysis} for every edge, fog, and cloud node. Based on the discussion on Subsection \ref{sec:criteria}, edge nodes require high data privacy as they are more frequently accessed than other nodes, and their resources are limited which in turn restricts them to applying other privacy and security measures (e.g., encryption, firewall, etc.) \cite{islam2021resource}. However, if the attacker can compromise the fog or central cloud he can exploit the aggregated data to conduct more devastating attacks. So, in this case, the fog and cloud nodes require better privacy than the edge node. 
\subsection{ Risk Factor Analysis}
\noindent In risk factor analysis, we consider the risk of disclosure and re-identification of individual data points. The risk factor is high when the data privacy level is low. More specifically, if the noise of the differential privacy mechanism is low then the attacker can easily re-identify individual data points by conducting multiple aggregation queries and averaging out the results. This correlation is also reflected in our empirical evaluation outlined in the discussed part of this paper.
    
\begin{figure*}[b!]
\centering
\includegraphics[width=0.99\linewidth]{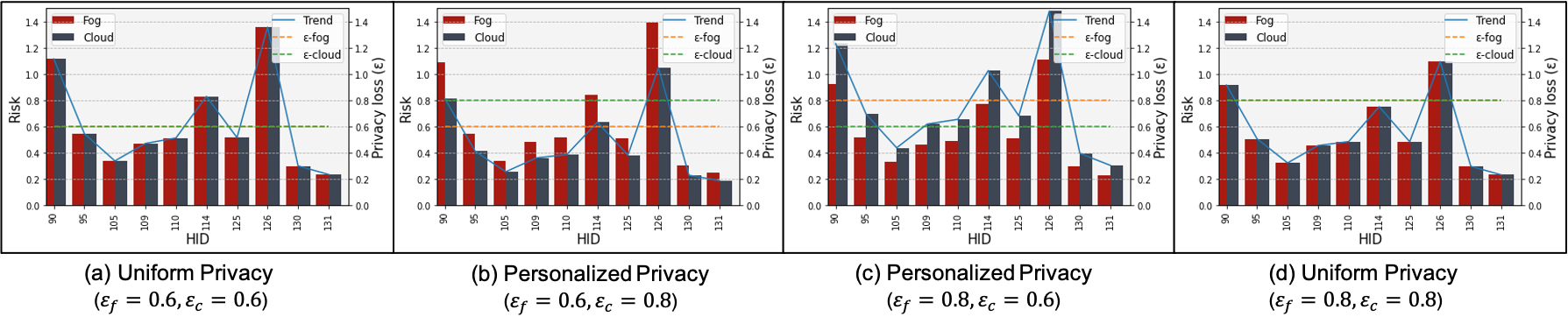} 
\caption{Comparative analysis between privacy loss and disclosure risk of private information for uniform and personalized privacy.} 
\label{fig: privacy_cost}
\vspace{-10pt}
\end{figure*}

\subsection{Simulation and Discussion}

\subsubsection{DP Evaluation}
\noindent In this section, we conducted our experimental analysis using a micro-grid data-set of 443 homes collected from UMASS Repository \cite{zhu2011case}. The dataset contains minute-level data of a particular day (24 hours) for these 443 homes. Fig. \ref{fig: Personalized_DP} (a) illustrates the corresponding features of the dataset (e.g., consumption, home id, and timestamp). We can see that during the busy hours starting from 8 a.m to 6 p.m the consumption is significantly higher than other hours. Moreover, approximately, the first 100 homes consume higher energy than the rest of the homes. We considered these minute-level data as the granular data and the homes as the edge nodes. Then, we observed the impact of differential privacy over all the granular data of these edge nodes (Fig. \ref{fig: Personalized_DP} (b). It can be inferred that higher privacy loss (alternatively, lower noise) yields better data utility. That means if the privacy loss is increased (e.g., $\varepsilon$  increased gradually from 0.1 to 0.4 in the figure) the privatized value remains more close to the granular value. Furthermore, we investigated the probability distribution of the differentially private data as shown in fig \ref{fig: Personalized_DP} (c). The probability distribution function shows that the laplacian noise of the differential privacy mechanism is more clustered around the mean when we increase the privacy loss (i.e., when $\varepsilon$ = 0.4). Contrarily, the noise spreads out more and more if we continue to decrease the privacy loss ($\varepsilon$). Here, the PDF of the fatter tailed curve ($\varepsilon$ = 0.1  in Fig. \ref{fig: Personalized_DP} (c)) provides more noise and thus ensures better privacy.

\subsubsection{Comparative Analysis between Personalized and Uniform Privacy}
To achieve a comprehensive understanding of the importance of applying PDP over UDP, we selected a set of two different privacy levels for both the fog and cloud aggregation which is $\varepsilon= 0.6$ and $\varepsilon = 0.8$. Data aggregation is not required in the edge nodes; rather aggregation function is performed while transferring the data from the edge node to the fog and fog to the cloud. In UDP setting, two cases are demonstrated where both the fog and cloud aggregation adopts the same privacy levels at a time (i.e., case $1$: ($\varepsilon_f, \varepsilon_c$) = \{$0.6, 0.6$\}; case $4$: ($\varepsilon_f, \varepsilon_c$) = \{$0.8, 0.8$\}). However, in the PDP set up we considered the same privacy level but in an alternative manner (i.e., case $2$: ($\varepsilon_f, \varepsilon_c$) = \{$0.6, 0.8$\}; case $3$: ($\varepsilon_f, \varepsilon_c$) = \{$0.8, 0.6$\}). Fig. \ref{fig: loss_DP} (a) and (b) depict the loss distribution of these four cases for uniform and PDP accordingly while Fig. \ref{fig: loss_DP} (c) presents the mean and standard deviation of these loss distributions. Generally, the higher standard deviation yields better privacy since the noise is distributed widely from the mean value. Also, the higher mean yields to higher loss (alternatively, lower data utility). So, for the case $1$, where both the standard deviation and mean are high, privacy will be highest but the data utility will be lowest. In contrast, for the case $4$, the privacy will be the lowest, and data utility will be the highest. So, if we consider uniform privacy, it may offer excessive privacy control to a subset of nodes while applying insufficient protection to another subset.

However, if we consider the PDP setting, we can achieve optimized data privacy and security for this we need to consider cases $2$ and $3$. In the application, where data privacy is more desirable than the data utility case $2$ will be preferred over case $3$ and vice versa. We have considered privacy loss $0.6$ and $0.8$ in our experiment but different sets of privacy loss can also be opted. Nevertheless, the distribution follows the same trend as Fig. \ref{fig: loss_DP} for any set of privacy loss.

\subsubsection{Correlation among Privacy Loss, Data Utility and Data Disclosure Risk }

We outlined the data disclosure or re-identification risk for the same four cases under uniform and PDP settings as shown in Fig. \ref{fig: privacy_cost}. The correlation among the privacy loss, data utility, and data disclosure risk shows that risk and utility decrease along with the increment of privacy loss ($\varepsilon$) in the UDP settings. For instance, in case of house id (HID) $114$ if we increase the privacy loss from $0.6$ to $0.8$, the risk decreases approximately from 0.81  (Fig. \ref{fig: privacy_cost}(a)) to $0.79$ (Fig. \ref{fig: privacy_cost}(d)) for both the fog and cloud aggregation. However, if we consider the PDP for the same house, the minimum attainable risks can be $0.61$ (for cloud aggregation, Fig. \ref{fig: privacy_cost}(b)) and $0.79$ (for fog aggregation, Fig. \ref{fig: privacy_cost}(c))). If the CPS analysts or customers pull the data from the cloud node, then the case $2$ (i.e., $\varepsilon_f, \varepsilon_c$ = \{$0.6$, $0.8$\}, Fig. \ref{fig: privacy_cost}(b)) is more preferable over the rest three cases since the cloud risk factor is the lowest for this case. 

Moreover, a little increment in the privacy loss ($\varepsilon$) of fog aggregation ($\varepsilon_f$) yields to larger risk increment in the cloud aggregation. For example, for the same house id (i.e., $114$), increasing $\varepsilon_f$ from $0.6$ to $0.8$ increases the cloud aggregation risk from $0.81$ to $1.01$ (Fig. \ref{fig: privacy_cost}(a) and Fig. \ref{fig: privacy_cost}(c)). Although the UDP decreases the disclosure risk factor by increasing the fog and cloud privacy losses ($\varepsilon_f, \varepsilon_c$), it does not consider the data utility in the process. In contrast, a PDP mechanism tunes both the fog and cloud privacy loss levels to the optimized values ($\varepsilon_f^*, \varepsilon_c^*$) to attain optimal data privacy and data utility with optimal disclosure risk.  

\section{Conclusion and Future Work}\label{sec:Conclusion}
\noindent This paper demonstrates a comprehensive vulnerability characterization and privacy quantification as well as a protection model to deal against the security and privacy-related issues in CPS during data generation, transmission, and collection. The proposed SVPL model helps to identify security and privacy vulnerabilities of the CPS system accurately by building a standard vulnerability profile considering each data provider's privacy loss scenario. Furthermore, a customized, PDP model is also introduced to improve the data utility by reducing data disclosure risk while assuring high-level data privacy protection for the CPS. Finally, our experimental analysis compares proposed personalized differential privacy (PDP) with uniform differential privacy and proves that this PDP model ensures better data privacy protection while maintaining the data utility and minimizing the risk of data disclosure. 

In the future, we plan to demonstrate a real-life privacy characterization model using IEEE bus systems and formulating different node attack strategies to show the SVPL model's effectiveness, which is currently out of this paper's scope.

\bibliographystyle{IEEEtran}
\bibliography{bibliography.bib}

\end{document}